# Gate-tunable high mobility remote-doped InSb/In$_{1-x}$Al$_x$Sb quantum well heterostructures

Wei Yi,[1,*] Andrey A. Kiselev,[1] Jacob Thorp,[1] Ramsey Noah,[1] Binh-Minh Nguyen,[1] Steven Bui,[1] Rajesh D. Rajavel,[1] Tahir Hussain,[1] Mark Gyure,[1] Philip Kratz,[2] Qi Qian,[3] Michael J. Manfra,[3] Vlad S. Pribiag,[4,†] Leo P. Kouwenhoven,[4] Charles M. Marcus,[5] Marko Sokolich[1,*]

1. HRL Laboratories, 3011 Malibu Canyon Rd, Malibu, CA 90265

2. Department of Applied Physics, Stanford University, Stanford, California 94305, USA

3. Department of Physics and Astronomy, Purdue University, 525 Northwestern Ave, West Lafayette, IN 47907

4. Kavli Institute of Nanoscience, Delft University of Technology, 2600 GA Delft, The Netherlands

5. Center for Quantum Devices, Niels Bohr Institute, University of Copenhagen, 2100 Copenhagen, Denmark

Abstract:

Gate-tunable high-mobility InSb/In$_{1-x}$Al$_x$Sb quantum wells (QWs) grown on GaAs substrates are reported. The QW two-dimensional electron gas (2DEG) channel mobility in excess of 200,000 cm$^2$/Vs is measured at T=1.8K. In asymmetrically remote-doped samples with an HfO$_2$ gate dielectric formed by atomic layer deposition, parallel conduction is eliminated and complete 2DEG channel depletion is reached with minimal hysteresis in gate bias response of the 2DEG electron density. The integer quantum Hall effect with Landau level filling factor down to 1 is observed. A high-transparency non-alloyed Ohmic contact to the 2DEG with contact resistance below 1Ω·mm is achieved at 1.8K.

*Corresponding authors: wyi@hrl.com and MSokolich@hrl.com

†Current address: School of Physics and Astronomy, University of Minnesota, Minneapolis, Minnesota 55455, USA

InSb has the smallest electron effective mass ($m^* = 0.014 m_e$) and the largest known room-temperature bulk electron mobility of 78,000 cm$^2$/Vs of any semiconductor, making it appealing as a channel material in heterostructure transistors (HEMT, HBT) that can operate at higher frequency and lower voltage than GaAs and InP based devices [1]. The strong spin-orbit interaction and a giant g factor (-51 in bulk InSb) makes InSb an interesting candidate for potential exploitation in spintronics [2] or topological quantum computing using non-Abelian Majorana fermions [3,4]. In fact, signatures of Majorana fermions were first reported in solid state systems using a gated InSb nanowire-superconductor hybrid device [5-7]. However, integration of bottom-up nanowires remains a difficult task. To scale to more complicated structures for topological qubits, an alternative approach is to fabricate InSb nanowires and "T" junctions by top-down techniques starting from planar epitaxial heterostructures. Electron mobility greater than 100,000 cm$^2$/Vs and a corresponding electron mean-free path in the sub-micrometer range are achievable. Although InSb/InAlSb QWs with 2DEG mobility in excess of 200,000 cm$^2$/Vs have been developed [8,9], they are often plagued with parallel conduction channels which may be difficult to deplete. Gate tunability of the 2DEG density to full depletion is needed in split-gate quantum structures. However, making gated devices for InSb-based materials has been a major challenge. Schottky barrier gates to such narrow-gap materials are leaky due to the low barrier height [10]. For metal-oxide-semiconductor (MOS) structures, previous studies found difficulties in growing high-quality gate dielectrics with appropriate interface properties to achieve depletion without hole accumulation or large hysteresis [11-14]. So far only one work claimed complete 2DEG channel depletion on InSb MOS device, with the 2DEG mobility reaching ~70,000 cm$^2$/Vs (at a 2DEG density of ~3.3x10$^{11}$ cm$^{-2}$). However, no data in the pinch-off regime was reported [14]. In this letter, we show that all the aforementioned challenges can be overcome with carefully designed epitaxial growth and device processing.

Previous studies showed that for remote-doped InSb/InAlSb QWs grown on lattice-mismatched GaAs substrates with typical rms surface roughness of the order of a few monolayers, the contribution of interface roughness scattering has a strong dependence on the well width and is negligible for a QW thickness of 30nm or greater. The low-temperature InSb 2DEG mobility is limited by remote ionized impurity (RII) scattering originating from ionized dopants (e.g. Te or Si) in the $\delta$-doping layer and charged defects in the buffer layer [9]. Screening by electrons in the QW and the $\delta$-doping plane both improve the mobility. However, electrons in the $\delta$-doping plane and higher subbands in the QW act as parallel conduction channels and are undesirable for certain device operations that require a single electron channel, e.g. the quantum Hall effect (QHE). Magnetotransport of a Hall device is well suited to diagnose the parallel conduction. In the absence of parallel conduction, a clean QHE with the longitudinal resistance showing zero-resistance states and quantized plateaus in Hall resistance should be observed. Otherwise the magnetoresistance oscillations are superimposed on a background of parallel conduction and never become zero (see sup. materials and Ref. [15]). Due to the rather low conduction band offsets in InSb/InAlSb QWs, the maximum 2DEG density that can be achieved without introducing parallel conduction is typically less than ~4x10$^{11}$ cm$^{-2}$ [16]. Inserting multiple doping layers with a goal of increasing the 2DEG density can produce undesirable parallel conduction channels. We have prepared and measured both symmetrically doped InSb QWs (with n-type doping layers at both above and below the QW) and asymmetrically doped QWs (i.e. with only n-type doping layer(s) above the QW). We found that the 2DEG density in a symmetrically doped QW is nearly doubled as compared with an asymmetrically doped QW. However, the parallel conduction in the bottom doping layer cannot be eliminated without first depleting the 2DEG channel. In contrast, asymmetrically doped samples do not have such issues. Therefore in this work we focus on results from asymmetrically doped samples. We report that by optimizing the design, epitaxial growth and fabrication process, an asymmetrically doped

InSb/InAlSb QW can achieve both 2DEG mobility higher than 200,000 cm$^2$/Vs and a nearly hysteresis-free gate bias response to full depletion.

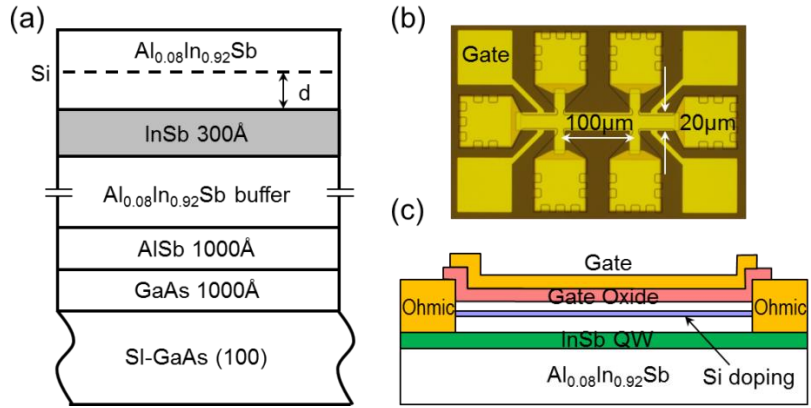

Fig1- (a) Schematic diagram of the heterostructure growth, (b) photograph of a gated Hall bar device, (c) schematic cross section of Hall bar device.

Fig. 1 shows the schematic diagram of the heterostructure growth and gated Hall device design used in this work. Samples were grown in a Varian Gen II molecular beam epitaxy (MBE) system equipped with valved crackers for As, Sb and effusion cells for Ga, In and Al. The InSb/InAlSb single QW heterostructure (Fig. 1a) was grown on a semi-insulating GaAs (100) substrate, on top of a GaAs buffer, an AlSb nucleation layer and a 4 micrometer thick strain-relaxed Al$_x$In$_{1-x}$Sb buffer layer (x=8%). The InSb quantum well thickness (30nm), the Si δ-doping level (5x10$^{11}$cm$^{-2}$-1x10$^{12}$ cm$^{-2}$) and the distance *d* to the QW (20nm) were selected after several rounds of test growths and Schrödinger-Poisson (SP) simulations to achieve high 2DEG mobility. The surface depth of the Si δ-doping was in the range of 20-50nm (20nm used for the data presented). Transmission electron microscopy (TEM) of the sample cross sections (Fig. S1 in sup. materials) showed that the defective region extends to about 3μm into the In$_{1-x}$Al$_x$Sb buffer layer and the buffer region beyond that thickness has relatively low density of dislocations. To inspect the effect of buffer thickness, we grew several samples with identical layer structures and growth conditions, except that the buffer thickness was changed to either 1μm or 6μm. For 1μm-buffer samples, the low-temperature 2DEG mobilities dropped from ~200,000 cm$^2$/Vs to ~60,000 cm$^2$/Vs, whereas for 6µm-buffer samples the mobility improvement over 4µm-buffer samples was marginal. The rms surface roughness of the as-grown sample surface is 1.3nm for 1µm-buffer samples and 1.9nm for 4µm-buffer samples over an area of 10µm x 10µm (for more details about the material growth, see sup. materials).

Processing of narrow-gap InSb-based materials poses another major challenge. At elevated process temperatures, several mechanisms can deteriorate the device characteristics, including but not limited to, interface roughening, preferential evaporation of Sb atoms on exposed surfaces, and diffusion of In and Sb into gate dielectrics [17]. As a precaution, we have kept the process thermal budget conservatively at or below 180°C. For high-frequency transistors and superconductor-semiconductor hybrid quantum devices, a low resistance Ohmic contact to the InSb 2DEG is desired for various reasons. Conventional Ohmic contacts on the top surface by eutectic alloys such as AuGe/Ni or AuGe/Pt [18] require a post-deposition anneal, which may cause material damage and surface leakage. Instead, we took the approach of etching to the InSb quantum well using an Ar ion etch with secondary ion end-point detection followed by non-alloyed Ti/Au evaporation (500Å Ti/3000Å Au). The contact resistance is characterized by transmission line method and is found to be less than 1Ω·mm at both room temperature and low temperatures (see sup. materials).

For fabricating gated MOS devices, a major challenge is growing high-quality gate dielectrics on (Al)InSb with low gate leakage. It was reported that the typical interface trap density $D_{it}$ for Al$_2$O$_3$ gate oxide by atomic layer deposition (ALD) was in the range of 1.1x10$^{12}$ cm$^{-2}$eV$^{-1}$ to 1.7x10$^{13}$ cm$^{-2}$eV$^{-1}$, and the gate depletion and hysteresis was improved by replacing the InSb surface layer with a larger bandgap Al$_{0.1}$In$_{0.9}$Sb [14]. ALD HfO$_2$ was shown to be a superior diffusion barrier than ALD Al$_2$O$_3$ for InSb [17]. Another advantage of ALD HfO$_2$ is its much larger dielectric constant (κ ~ 18-21) than ALD Al$_2$O$_3$ (κ ~7-9).

We deposited a 20-40nm high-κ ALD HfO$_2$ thin film as the gate dielectric using remote plasma enhanced ALD (PEALD) method at a low substrate temperature of 175°C. PEALD is expected to reduce impurities and increase the film density as compared with thermal ALD process. A dielectric breakdown field of ~3MV/cm was measured at room temperature.

For magnetotransport characterizations, materials were processed into 6-lead Hall bars using standard photolithography techniques. 200 μm long, 20 μm wide Hall bar mesas with Hall voltage lead spacing of 100 μm were wet etched using citric-acid based solution [19](see sup. materials for details). A top view micrograph of an actual device is shown in Fig. 1(b) with a schematic cross section shown in Fig. 1(c). Processed devices were tested at 1.7-1.8 K in a Lakeshore 9709A Hall measurement system equipped with a 9T magnet using AC lock-in technique (frequency at 9Hz, time constant at 30 ms). The Hall bar's drain/source were connected in series with a 25 kΩ load resistor under a 10 mV AC bias, which restricts the channel current to below 400nA to avoid localized Joule heating. Measured gate leakage was at or below the noise floor of the instrument (~20pA) for all DC biases applied (up to ±5V).

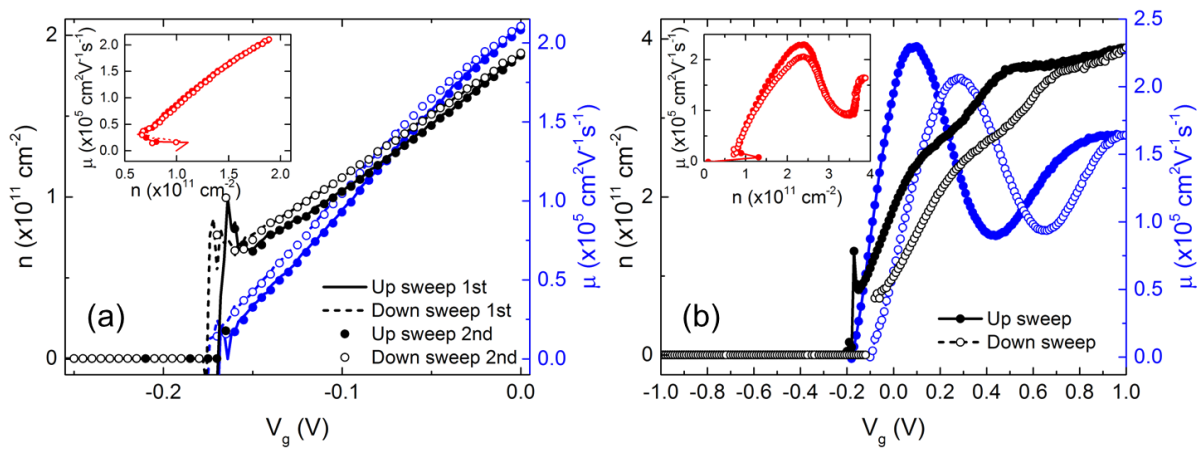

Fig2- Electron density and mobility of the InSb 2DEG as a function of gate bias $V_g$ measured at B = 0.1 T (low field). (a) Depletion-mode gate bias double sweeps (0V to -0.3V then back to 0V, repeated twice).

(b) wide-range bipolar gate bias double sweeps (-1V to 1V then back to -1V) of the same device. Inset shows the mobility vs. electron density.

The field dependent magnetoresistance $R_{xx}$ and Hall resistance $R_{xy}$ allow for extraction of carrier mobility and sheet charge density. The linearity of $R_{xy}$ vs. B (not shown) suggests that single B field measurement is adequate. Fig. 2 shows electron density and mobility of the InSb 2DEG as a function of gate bias, $V_g$, measured at B = 0.1 T (low field). In Fig. 2 (a), repeated negative Vg double sweeps (0V to -0.3 V then back to 0V) were applied to a Hall bar to deplete/replenish the 2DEG channel. A small hysteresis of ~20mV was seen in the initial gate bias sweep after the Hall device was cooled to 1.8 K with all the electrical contacts grounded (data not shown). In all subsequent sweeps, the hysteresis was reduced to only ~10mV and remained the same irrespective of the voltage sweep rate. The mobility vs sheet charge density (Fig. 2(a) inset) follows a linear dependence, which suggests that carrier scattering is dominated by the interplay of remote dopants and background impurities. The 2DEG channel is completely pinched off at a small negative gate bias of $V_P$ = -0.17 V ($V_P$ varies slightly from device to device and is in the -0.1V to -0.25V range). The channel pinch-off always occurs abruptly when the 2DEG electron density drops below certain thresholds (2 to $6 \times 10^{10}$ cm$^{-2}$, varies with device), suggesting a percolation type of behavior with a minimum electron density for metal-insulator transition to occur [20]. No sign of hole accumulation was observed in samples using a top Si δ-doping level of $1 \times 10^{12}$ cm$^{-2}$. However, hole accumulation was observed in samples using a lower Si doping level of $5 \times 10^{11}$ cm$^{-2}$ (see sup. materials). This suggests that a potential barrier in the valence band produced by the δ-doping layer is important in preventing hole accumulation towards the surface, which requires further investigation.

The situation is different when a positive gate bias sweep is used. In Fig. 2(b), bipolar wide-range gate bias double sweeps (-1V to 1V then back to -1V) were applied to the same Hall device. Electron density in the 2DEG channel keeps increasing at higher positive biases, while mobility peaks at 230,000 cm$^2$/Vs at $V_g$ = 0.1V then drops at higher gate biases, with a corresponding decrease in the slope of n-$V_g$. Both can be explained by the onset of filling into the 2nd subband which has a lower mobility due to hybridization with the δ-doping layer, as shown by SP simulation (see Fig. 3). The 2DEG channel turn-on threshold shifts to a larger value after positive $V_g$ sweeps and a much larger hysteresis is produced accordingly. However the μ vs. n relationship is still consistent with low hysteresis (Fig. 2b inset).

The slope of n-$V_g$ can be used to estimate the interface trap density using an equivalent circuit capacitor model of

$$dn/dV_g = C_{tot}/e[1 + C_{it}(C_{SC} + C_{2D})/(C_{SC}C_{2D})]$$

in absence of parallel conduction and hole accumulation [14]. Here $C_{it} = e^2 D_{it}$ is the capacitance associated with interface traps. $C_{ox}$ and $C_{sc}$ are the capacitance per unit area of the HfO$_2$ (dielectric constant = 18.5) and the semiconductor layers above the QW (dielectric constant = 16.4), $C_{2D} = e^2 m^*/\pi\hbar^2$ is the quantum capacitance of the 2DEG[21], $C_{tot}$ is the total capacitance of $C_{ox}$, $C_{sc}$, $C_{it}$, and $C_{2D}$. In negative $V_g$ sweeps (Fig. 2(a)), the $D_{it}$ value of 7.5x10$^{11}$ cm$^{-2}$eV$^{-1}$ is estimated from the n-$V_g$ slope of 8.4x10$^{11}$ cm$^{-2}$/V. This $D_{it}$ level in depletion-mode is close to a reported mean $D_{it}$ value of 8x10$^{11}$ cm$^{-2}$eV$^{-1}$ in HfO$_2$/InSb MOS capacitors [17], and is lower than a reported $D_{it}$ value of 1.1x10$^{12}$ cm$^{-2}$eV$^{-1}$ in Al$_2$O$_3$/InSb devices [14]. In positive $V_g$ sweeps (Fig. 2(b)), $D_{it}$ levels of 1.5x10$^{12}$ cm$^{-2}$eV$^{-1}$ and 3.3x10$^{12}$ cm$^{-2}$eV$^{-1}$ are estimated from the n-$V_g$ slopes of 7.3x10$^{11}$ cm$^{-2}$/V and 5.5x10$^{11}$ cm$^{-2}$/V for up and down sweeps, respectively. The large increase in $D_{it}$ under positive gate bias sweeps indicates filling of empty interface traps towards the conduction band as electrons tunnel from the QW into the surface under positive $V_g$. This adds negative interface charges and shifts the 2DEG turn-on threshold higher. It is also noteworthy that our PEALD process consistently produced negative pinch-off thresholds. For devices with pinch-off voltage near -0.2V, the 2DEG mobility already reaches 200,000 cm$^2$/Vs at zero gate bias. Therefore a positive gate bias sweep is not needed in depletion mode operation and electron injection can be avoided. To overcome the issue of electron injection under positive gate bias, a higher Al-content barrier can be used above the QW to reduce the electron tunneling across the InSb/InAlSb interface.

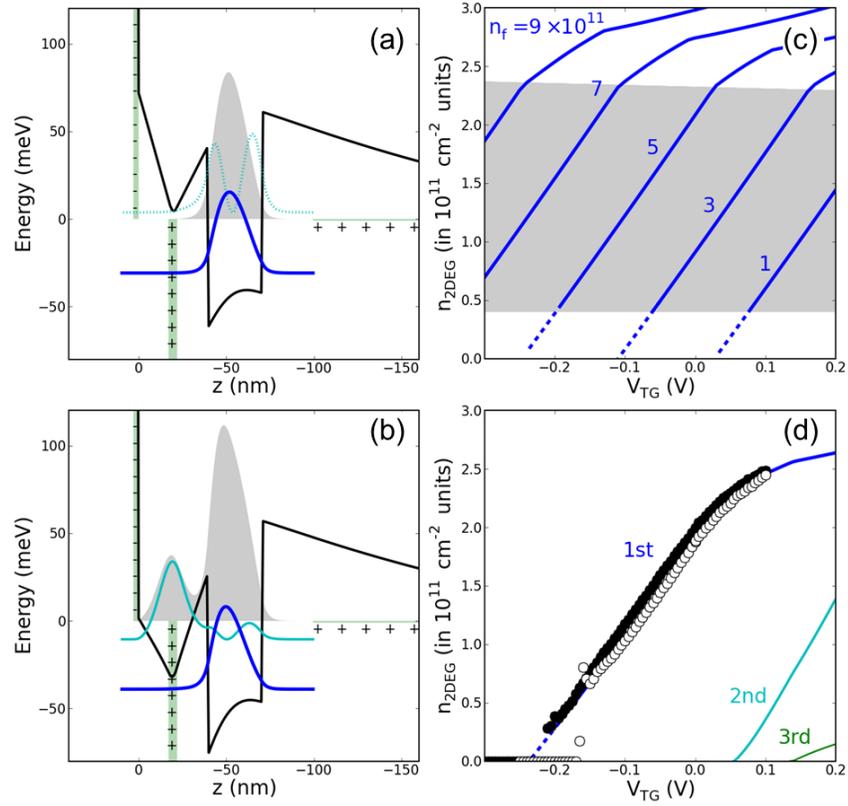

Fig3- Self-consistent Schrödinger-Poisson simulations of the top-doped structure. (a) Conduction band diagram (CBD) at $V_g = 0$ when only a single channel (the QW first subband) is populated. Mobile charge density is shown as shaded area. Position of the intentional top Si doping ($n_f = 5 \times 10^{11}$ cm$^{-2}$) along with that of the weakly $V_g$-dependent dielectric interface charge density (for Dit = $7.5 \times 10^{11}$ cm$^{-2}$eV$^{-1}$, total $n_{Surface} \sim -8 \times 10^{11}$ cm$^{-2}$) and a weak intrinsic bulk doping ($n_{Bulk} = 1 \times 10^{15}$ cm$^{-3}$) are depicted schematically. Envelope function density for electron states at the bottom of the first and second subbands are shown by solid (blue) and dotted (cyan) lines, respectively. (b) CBD at $V_g = 0.15$ V, wherein two channels — one in a QW and one in the doping layer — are populated appreciably. (c) First channel charge density as a function of the fixed charge density $n_f$ in the doping layer. Shaded region depicts the desired single channel regime. (d) Mobile charge density (channel-partitioned) vs top gate bias. Experimental data for two different devices are shown by solid and open circles.

Better understanding of the self-consistent steady-state charge distribution and formation of the mobile channels in the top-gate biased structure has been achieved by employing Schrödinger-Poisson simulations. Electron spatial quantization and Thomas-Fermi filling of the resulting 2D subbands are described in the effective mass approximation, accounting for the conduction band non-parabolicity (which is prominent in narrow-gap materials like InSb). Material parameters for InAlSb alloy-based heterostructure can be found in Ref. [22]. Subplots (a) and (b) of Fig. 3 provide examples of the conduction band profile at two distinct gate biases ($V_g$ = 0 and 0.15 V here, $n_f$ = 5x10$^{11}$ cm$^{-2}$), allowing substantial electron density (shown as shaded area) accumulation in only a single conducting channel or two parallel channels, respectively. Electrons in the first channel are always strongly localized in the QW (envelope function density at the bottom of the first subband is sketched by the blue line). Simulations suggest that electrons in the emerging second channel experience strong hybridization between states in the doping layer and the QW second subband (cyan line), and thus are expected to have lower mobility. The amount of fixed charge in the doping layer (due to different dose and/or fraction of activated dopants) is not fully known and would require extensive additional characterization. In Fig. 3(c) charge density in the first channel is shown as a function of the top gate bias for a set of $n_f$ values which is treated here as a free input parameter. Its main effect is a parallel shift of gate biases that would be required to bring the structure to a similar charge distribution (even identical when no mobile charges are present above the QW). Kinks mark sequential activation of additional channels. Dashes mark the approximate low threshold for metal insulator transition.

Results of the simulations at $n_f$ = 5x10$^{11}$ cm$^{-2}$ are compared with experimental n-$V_g$ Hall data for two devices in Fig. 3(d), where electron density is explicitly partitioned between channels. Below $V_g$ ~0.05 V only a single conducting channel is present and correspondence between simulation and experiment is good all the way down to the channel pinch-off observed at $n_{2DEG}$ of a few $10^{10}$ cm$^{-2}$.

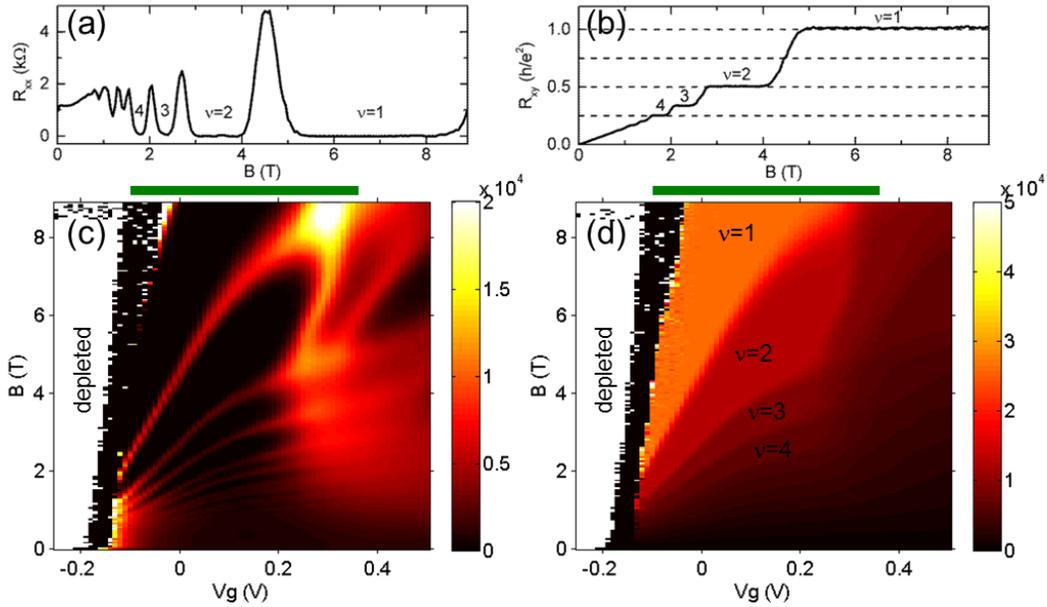

Fig4- field dependent (a) magneto resistance $R_{xx}$ and (b) Hall resistance $R_{xy}$ at $V_g$=0 V. 2D maps of (c) $R_{xx}$ and (d) $R_{xy}$ for all B fields (0 to 8.9 T) and gate biases (-0.25 to 0.5 V). Green bars show the regime wherein electron mobility of the InSb 2DEG is greater than 100,000 cm$^2$/Vs.

After establishing gate tunability of the 2DEG density to full depletion with minimal hysteresis, we performed Hall measurements at high B field to examine the existence of a parasitic conduction channel, which is a common issue for InSb based narrow-gap heterostructures. High B field sweeps manifest no sign of parallel conduction between the 2DEG channel pinch off and the mobility peak (at $V_g$ = 0.08 V) resulting in a clearly observable quantum Hall effect (QHE). The Hall resistance shows well-defined quantized plateaus with a Landau level filling factor down to 1 and correspondingly dissipation-less zero-resistance edge states in magnetoresistance, as visualized in Fig. 4 (a) to (d). For most of the QHE regime, the low-field mobility of 2DEG channel is greater than 100,000 cm$^2$/Vs (green bars in Fig. 4 (c) and (d)).

In conclusion, we have demonstrated efficient gate control and full depletion of high mobility InSb/InAlSb quantum wells grown on GaAs substrates. A high-κ PEALD HfO$_2$ dielectric forms an excellent interface with InAlSb and delivers nearly hysteresis-free gate response at very low operating voltages. Combined with a 2DEG mobility greater than 100,000 cm$^2$/Vs in most of the QHE regime and a low Ohmic contact resistance of less than 1Ω·mm, this development facilitates fabricating top-down InSb-based nanostructures suitable for spin and Majorana fermion based quantum computation schemes.

We would like to acknowledge Gunjana Sharma, Jack A. Crowell, Elias A. Flores, Ivan Alvarado-Rodriguez, Marcel M. Musni, Kasey C. Fisher, John S. Yeah, Adele E. Schmitz, Helen K. Fung, Fiona C. Ku and David H. Chow at HRL labs and Morten Kjaergaard, Henri Suominen, and Fabrizio Nichele at Center for Quantum Devices, Copenhagen, for technical support and fruitful discussions. The authors are grateful for support from Microsoft Station Q.

# Supplementary materials

Wei Yi,[1,*] Andrey A. Kiselev,[1] Jacob Thorp,[1] Ramsey Noah,[1] Binh-Minh Nguyen,[1] Steven Bui,[1] Rajesh D. Rajavel,[1] Tahir Hussain,[1] Mark Gyure,[1] Philip Kratz,[2] Qi Qian,[3] Michael J. Manfra,[3] Vlad S. Pribiag,[4,†] Leo P. Kouwenhoven,[4] Charles M. Marcus,[5] Marko Sokolich[1,*]

1. HRL Laboratories, 3011 Malibu Canyon Rd, Malibu, CA 90265

2. Department of Applied Physics, Stanford University, Stanford, California 94305, USA

3. Department of Physics and Astronomy, Purdue University, 525 Northwestern Ave, West Lafayette, IN 47907

4. Kavli Institute of Nanoscience, Delft University of Technology, 2600 GA Delft, The Netherlands

5. Center for Quantum Devices, Niels Bohr Institute, University of Copenhagen, 2100 Copenhagen, Denmark

*Corresponding authors: wyi@hrl.com and MSokolich@hrl.com

†Current address: School of Physics and Astronomy, University of Minnesota, Minneapolis, Minnesota 55455, USA

# Epitaxial Growth

A key component of the MBE growth is to calibrate the substrate temperature which is challenging due to the infrared absorption of the (Al)InSb material. We took advantage of the fact that the reflection high-energy electron diffraction (RHEED) pattern on the (Al)InSb surface transforms from a pseudo (1x3) to center (4x4) when the temperature drops across a transition temperature of 380°C [1]. As grown materials were characterized using high resolution x-ray diffraction (HRXRD) to check the crystallinity and evaluate the InAlSb composition. The XRD peaks match very well with simulation using PANalytical's X'Pert Epitaxy software and the specified layer thicknesses in the growth. The fitted Al content x = 8.3% in the $In_{1-x}Al_xSb$ buffer is very close to the targeted value of 8%. Using TEM, the InSb quantum well thickness was measured to be 31nm, which is close to the targeted value of 30nm.

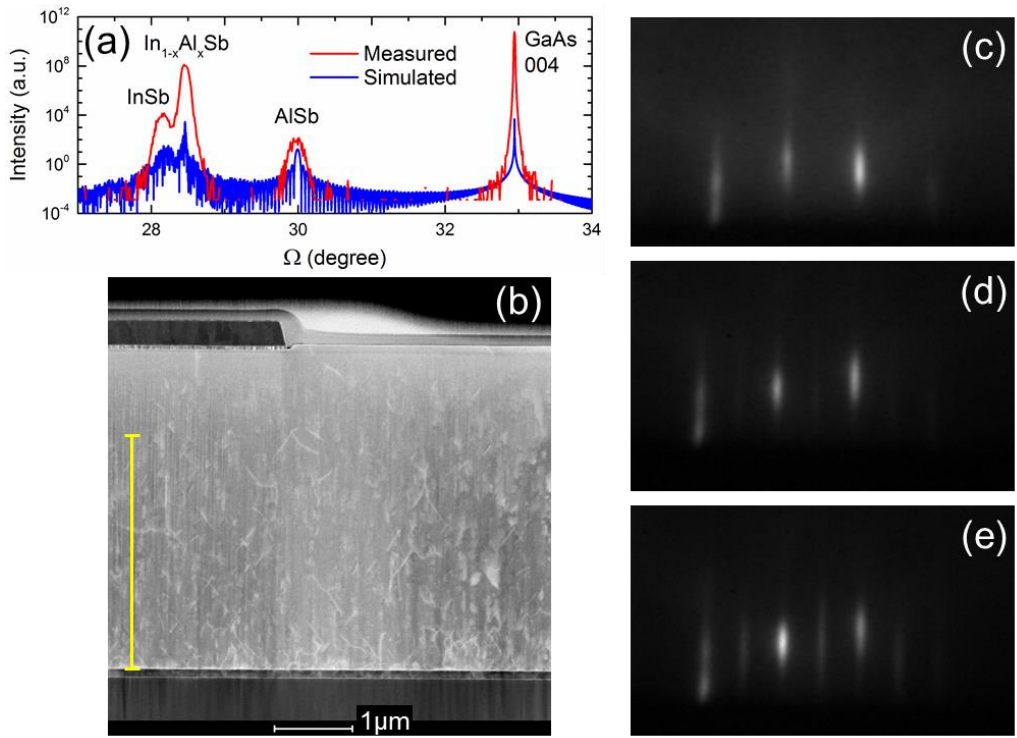

FigS1- (a) XRD and simulation of an $InSb/In_{1-x}Al_xSb$ (nominal x = 8%) heterostructure. The simulation was performed using the nominal layer thicknesses in the MBE growth and x as a free parameter. Best fit was found at x = 8.3%. (b) dark-field (DF) TEM image of the cross section of an InSb/ In$_{1-x}$Al$_x$Sb Hall bar showing much less dislocation density after grown 3um In$_{1-x}$Al$_x$Sb buffer layer (yellow line). RHEED patterns on the In$_{1-x}$Al$_x$Sb buffer at substrate temperature 495°C (C), 419°C (d), and 380°C (e), respectively illustrating a transition from pseudo (1x3) in (c) to center (4x4) in (e) across the transition temperature at 380°C.

**Wet Etch**

Hall bar mesas were wet etched using an aqueous citric-acid based solution that combines citric acid ($C_6H_8O_7$) and phosphoric acid ($H_3PO_4$) with hydrogen peroxide ($H_2O_2$). Etch of (Al)InSb is triggered by the oxidizing agent $H_2O_2$, and citric acid acts as a complexing agent to assist in the dissolution of antimony. On the (Al)InSb (001) surface, this wet etching is isotropic in nature and produced curved mesa sidewall profiles (data not shown). We have experimented with etch solutions with and without phosphoric acid ($H_3PO_4$) as a component. Solutions without $H_3PO_4$ showed relatively slow (Al)InSb etching, with a maximal etch rate of ~7Å/sec in a high-concentration mixture of citric acid/$H_2O_2$ = 0.5M/0.5M and relatively poorer run-to-run reproducibility. We added $H_3PO_4$ with various molarities into the mixture and found that the (Al)InSb etch rate increases linearly with the $H_3PO_4$ molarity x in the solution and reaches ~45 Å/sec at x = 1M (see Fig. S2 (a) and (b)). The run-to-run reproducibility was also improved. Citric acid is considered a weak acid and the range of pH value it can achieve is limited. In the absence of stronger acid, dissolution of oxidized species becomes the rate-limiting factor. Phosphoric acid is a much stronger acid and acts as a pH adjusting agent that controls the etch rate in such scenarios. All the Hall bar mesas were etched in a solution of citric acid/ $H_3PO_4$/ $H_2O_2$ = 0.5M/1M/0.5M. We preformed surface roughness measurements by atomic force microscopy (AFM) before and after the citric acid etch, and found the surface roughness actually decreased slightly after the wet etch, from 1.9 nm to 1.6 nm. This suggests that citric acid-based InSb etch is homogeneous in nature.

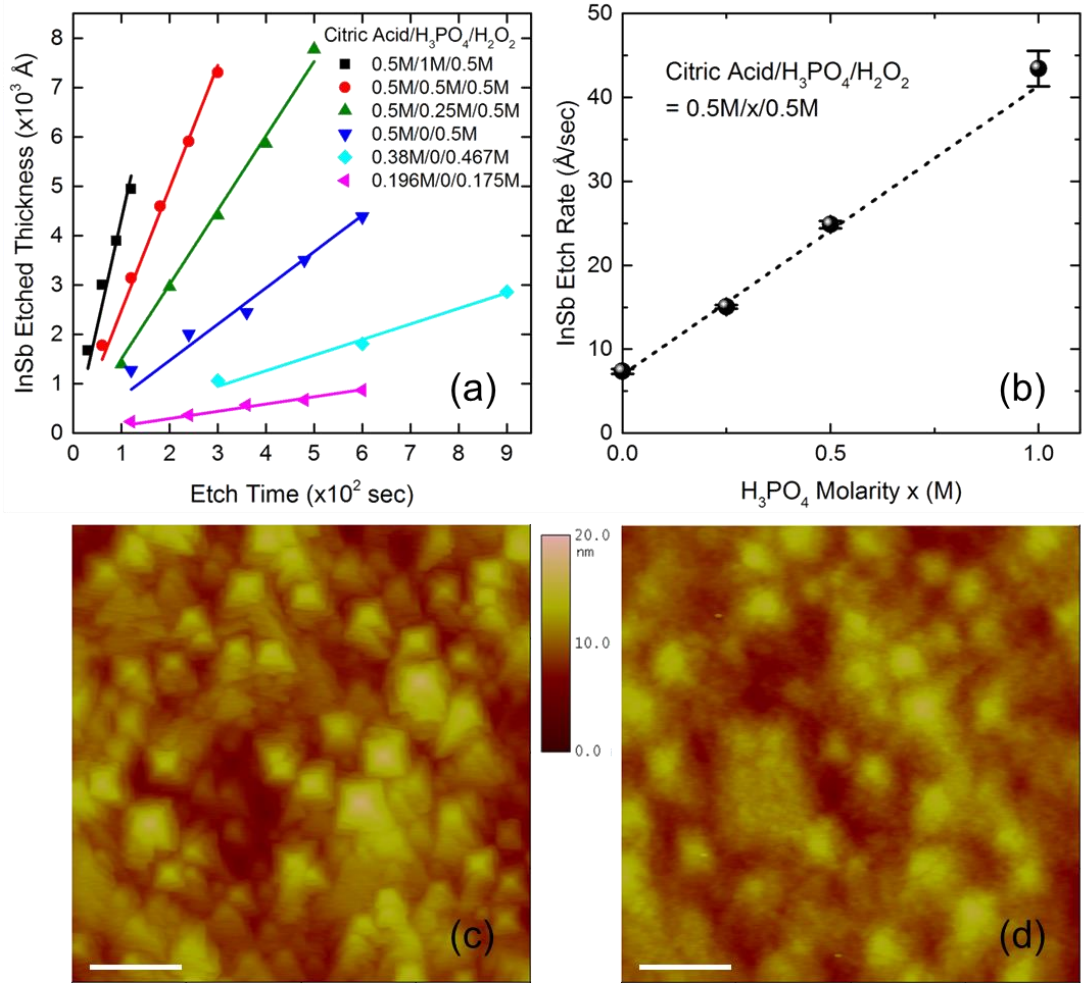

FigS2- (a) (Al)InSb etched thickness versus etch time in citric acid etch mixtures with various citric acid/$H_3PO_4$/$H_2O_2$ molarities as marked. (b) (Al)InSb etch rate versus phosphorous acid molarity x in citric acid etch mixtures of citric acid/$H_3PO_4$/$H_2O_2$ = 0.5M/x/0.5M. AFM images of a 10μm x 10μm area on the (Al)InSb surface before (c) and after (d) a 200 sec etch in a solution of citric acid/$H_3PO_4$/$H_2O_2$ = 0.5M/1M/0.5M (~0.9 μm removed). The rms roughness measured is 1.9 nm (c) and 1.6 nm (d), respectively. Scale bars are 2 μm.

**Ohmic contact**

Conventional alloyed Ohmic contacts for III-V materials requires a physical deposition (evaporation or sputtering) of eutectic metal stacks (e.g. AuGe/Ni) on the semiconductor surface, followed by a post-metal anneal (PMA) at above the eutectic transition temperature (361 °C for Au-Ge). For low-melting-point semiconductors such as InSb (melting point 527°C), the PMA process may be destructive and cause interfacial atomic interdiffusion and/or surface atom segregations which deteriorate certain device characteristics. Non-alloyed contacts, on the other hand, typically require etch back to a specific depth to achieve low contact resistance. However, precise control of the etch depth is challenging for InSb/InAlSb (Al < 30%) heterostructures that lack a good dry/wet etch selectivity. Chemical dry etch using optical end point detection methods (e.g. optical emission spectroscopy) does not have the sensitivity needed to detect the InSb/InAlSb heterointerface and can render poor etch depth reproducibility.

We adopted non-alloyed Ohmic contact utilizing Ar ion mill etch back equipped with a quadrupole mass spectrometer end point detector. A commercial Denton ion etch tool was used to develop the ion mill etch back method. The ion mill was performed at normal incident angle with no tilt or sample rotation. After ion mill, the sample was immediately transferred to an ebeam evaporator for metal deposition to minimize the impact of surface oxidization. The atomic sputtering rate is found to be proportional to the product of the Ar ion kinetic energy and the ion flux. By adjusting both the beam voltage and the beam current, etch rates in the range of 1.5-2.5Å/s suitable for shutter control of etch stop were achieved. The beam voltage/current used for 1.5 Å/s and 2.5 Å/s etch rates were 200 V/40 mA and 250 V/50 mA, respectively. The Ar gas flows for the source beam and the neutralizer were 5 sccm and 3 sccm, respectively. SIMS signals from the cation materials (Al and In) were monitored in real time throughout the ion mill. In calibration runs, the Al cation signal drops by ~90% as the InSb quantum well (QW) gets etched into. After the QW is completely etched, the Al count rate recovers to nearly its original level (data not shown). The drop in Al cation counts was used to stop the ion mill at the top interface of the InSb/InAlSb quantum well. We found that the Ar ion etch is non-selective for InSb or InAlSb (Al $\leq$ 14%) and the measured etch rates were essentially the same for samples capped with either InSb or InAlSb surface layers. The calculated coefficient of variation for the etch rates measured in 5 separate runs over a time period of six month was merely 2.9%, confirming an excellent run-to-run reproducibility.

The smoothness of ion etched surface was revealed by AFM roughness measurement (data not shown). The global roughness measured over an area of 20μm x20μm is less than 2 nm and is comparable to as-grown surfaces. The Ohmic contact interface is examined by cross-sectional TEM (Fig. S3). Fig. S3 (a) shows a high angle annular dark field (HAADF) scanning transmission electron microscopy (STEM) image that is capable of resolving the InSb quantum well by the Z contrast of different atomic weights for In and Al atoms. The ion etch time was calculated by the nominal top $In_{0.92}Al_{0.08}Sb$ barrier layer thickness with a target etch stop at the top surface of the InSb quantum well. The precision of the etch depth control is confirmed by the HAADF STEM image. Fig. S3 (b) shows a cross-sectional TEM image of the interface between Ti and the ion milled InAlSb layer. There is no clear evidence of an amorphous interfacial layer caused by Ar ion induced damage of the InAlSb lattice.

The Ohmic contact resistance is characterized at both room temperature and at low temperatures. Fig. S4 (a) shows a TLM (transmission line method) test structure containing non-alloyed Ti/Au Ohmic contacts at a series of different separations fabricated on an InSb/InAlSb mesa bar of 19μm width. In Fig. S4 (b), Ohmic contact pairs at three different gap distances (100μm, 50μm, 30μm) are measured at room temperature and 7.8 K. The linear-fit y-axis intercepts at zero gap distance are 70.3 Ohm and 86.2 Ohm at 300 K and 7.8 K, respectively. This resistance is composed of two contacts in series plus a small cable resistance (typically ~1 Ohm or less), which indicates a nearly temperature-independent contact resistance $R_c$ ~40 Ohm and a specific contact resistance of ~0.8 Ω·mm assuming only the front edge of the Ti/Au contact contributes to the current. This result is consistent with 2-terminal Hall bar channel resistance measurements at 1.8 K which found similar estimated specific contact resistance.

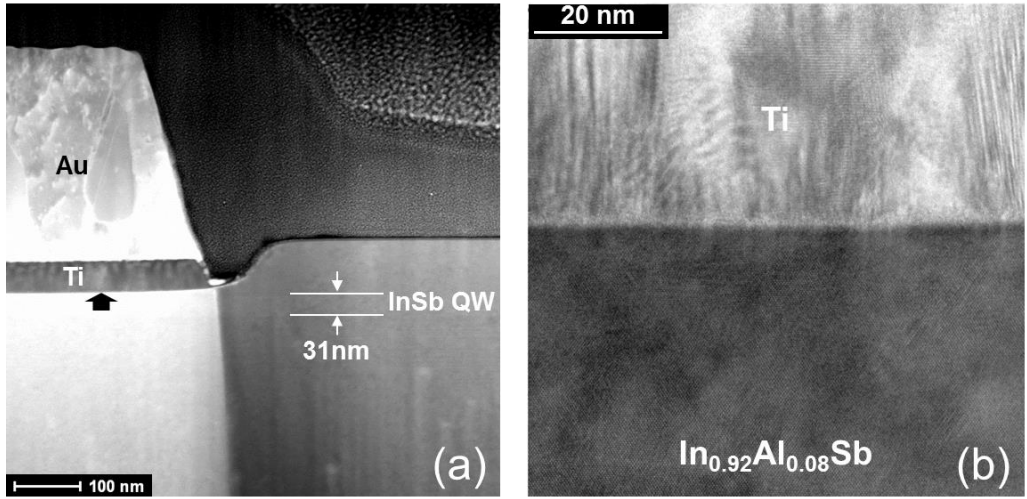

FigS3- (a) Cross-sectional Z-contrast image near the edge of a Ti/Au Ohmic contact to the InSb QW generated by high angle annular dark field (HAADF) scanning transmission electron microscopy (STEM) that shows the ion milled region of the InAlSb and the InSb QW (lighter contrast). Vertical line near the Ohmic contact edge is due to damage caused by focused ion beam TEM sample cut, (b) Cross-sectional TEM image of the interface between Ti and ion milled InAlSb layer.

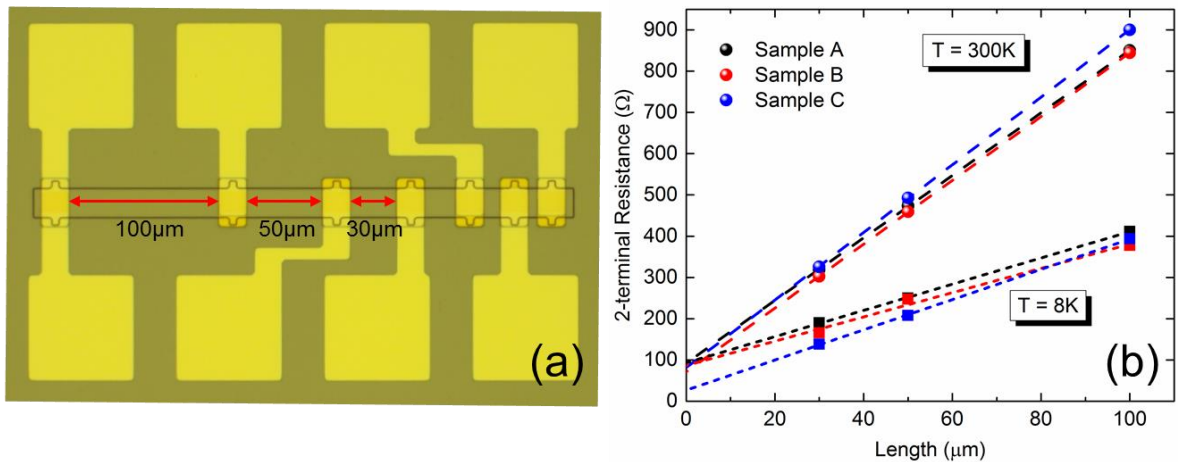

FigS4- (a) Top view micrograph of a TLM test structure showing Ohmic contacts at different separations made on the same mesa bar with 19μm width. (b) 2-terminal resistance versus the contact separation from 3 TLM devices (A, B and C) measured at room temperature and 8K. Dashed lines are linear fits.

**Gate dielectric**

Plasma-enhanced atomic layer deposition (PEALD) of the $HfO_2$ gate dielectric was performed in a commercial Cambridge Nanotech (now part of Ultratech) Fiji F200 ALD system equipped with a load lock and turbo molecular pump. Tetrakis(dimethylamido)hafnium(IV) (TDMAH, $Hf(N(CH_3)_2)_4$) and oxygen plasma were used as Hf and O source, respectively. TDMAH precursor was heated to 70 °C to maintain a vapor pressure of 1.9 Torr for efficient precursor delivery at a reactor pressure of ~200 mTorr. The sample chuck and reactor walls were heated to 175 °C during the deposition. Prior to ALD deposition, samples were first cleaned in acetone and isopropyl alcohol solvents for 1 min and $N_2$ blow dried, then a short-period (30 sec) remote O2 plasma descum was used to remove residual organic contaminates. Native oxides were removed in a two-step wet etch process. First a dilute buffered oxide etch (BOE (7:1) solution diluted to 2 %) dip of 15 sec followed by a deionized water rinse of 2 min, then a dilute ammonium hydroxide ($(NH_4)OH$ : water = 1 : 15) etch of 3 min followed by a deionized water rinse of 2 min and $N_2$ blow dry. Samples were transferred into the ALD system immediately after wet etch to avoid re-oxidization of the InSb surface. Before the ALD deposition cycles, 9 cycles of *in situ* surface pre-treatment were performed by applying a short period of $N_2$ plasma (100 Watt, 4 sec) followed by a Trimethylaluminum (TMA) pulse (40 msec) and 30 sec wait time to purge reaction products [2]. The $N_2$ plasma step was repeated once after the pre-treatment cycles. Each ALD deposition cycle was composed of an Hf half-cycle of a TDMAH pulse of 0.25 sec and a 15 sec purge to remove physisorbed precursor molecules, and an O half-cycle of $O_2$ plasma (300 Watt, 20 sec) followed by a 5 sec purge to remove reaction products. The $HfO_2$ growth rate per cycle (GPC) is 1.1 Å/cycle. The refractive index of the ALD HfO$_2$ films is 2.0-2.05. Dielectric strength of the ALD films was measured using a metal-oxide-semiconductor (MOS) structure on doped Si substrate. The oxide breakdown threshold field was found to be 2.85-2.95 MV/cm at room temperature.

## Additional Hall bar characteristics

Results from symmetrically remote-doped InSb/InAlSb quantum well heterostructures are summarized below. Fig. S5 shows the schematic diagram of the epitaxy layers. Fig. S6 (a) and (b) show electron density and mobility measured at a low B field (0.1 T). Fig. S6 (c) and (d) show B field-dependent magneto resistivity $\rho_{xx}$ and Hall resistance $R_{xy}$ at $V_g$=0 V, and $V_g$=0.9 V, respectively.

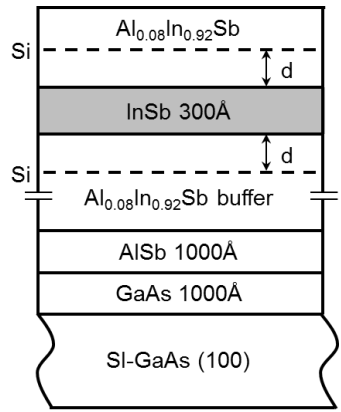

FigS5- Schematic diagram of a symmetrically remote-doped InSb/InAlSb heterostructure. The Distance $d$ between the Si $\delta$-doping and the QW was 30nm for this sample. The surface depth of the Si $\delta$-doping was 50nm.

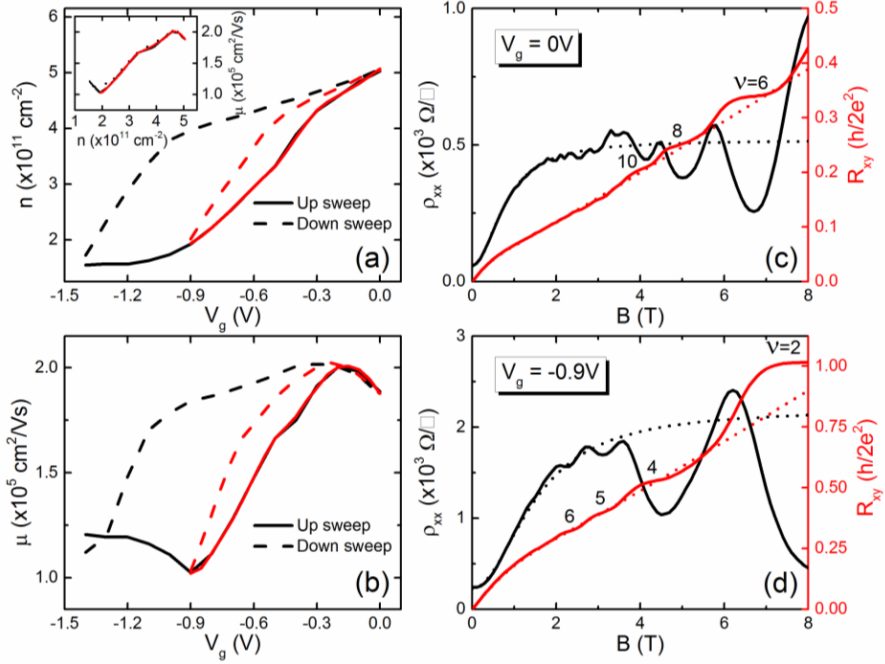

FigS6- (a) Electron density and (b) mobility of the InSb 2DEG as a function of gate bias $V_g$ measured at B = 0.1 T (low field) for the symmetrically doped sample. Solid lines are Up sweeps (start from 0V), dashed lines are Down sweeps (back to 0V). Inset shows the mobility vs. electron density. Field-dependent magneto resistivity $\rho_{xx}$ and Hall resistance $R_{xy}$ at $V_g$=0 V (c), and $V_g$=-0.9 V (d). Solid lines are measured data, dotted lines are least-squares fits by the two-carrier model (for data in B = 0-3.3 T). The fitted parameters for the 2DEG channel ($n_1$, $\mu_1$) and the parasitic channel ($n_2$, $\mu_2$) are: $n_1$=4.91x10$^{11}$ cm$^{-2}$, $\mu_1$=214,410 cm$^2$/Vs; $n_2$=5.13x10$^{11}$ cm$^{-2}$, $\mu_2$=6306 cm$^2$/Vs at $V_g$=0 V; and $n_1$=2.19x10$^{11}$ cm$^{-2}$, $\mu_1$=122,335 cm$^2$/Vs; $n_2$=2.25x10$^{11}$ cm$^{-2}$, $\mu_2$=3331 cm$^2$/Vs at $V_g$=-0.9 V. The ratio of electron densities between the 2DEG and the parasitic channel is ~0.96 and does not change noticeably with $V_g$.

The two-carrier model used to fit the magneto- and Hall resistivity data in Fig. S6 (c) and (d) is:

$$\rho_{xx} = \frac{\sigma_{xx}}{\sigma_{xx}^2 + \sigma_{xy}^2}$$

$$\sigma_{xx} = \sum_i \frac{n_i e \mu_i}{1 + (\mu_i B)^2}$$

$$\rho_{xy} = \frac{\sigma_{xy}}{\sigma_{xx}^2 + \sigma_{xy}^2}$$

$$\sigma_{xy} = \sum_i \frac{n_i e \mu_i^2 B}{1 + (\mu_i B)^2}$$